\documentclass[12pt]{article}
\usepackage{amsmath,epsfig,graphicx,verbatim,subfigure} 
\include{epsf}
\usepackage{cite}

\input{epsf} 
\providecommand{\href}[2]{#2}

\newcommand\as{\alpha_{\mathrm{S}}} 
\newcommand\f[2]{\frac{#1}{#2}} 
 
\def\to{\rightarrow}
\def\ito{\leftarrow} 
\def\nn{\nonumber}

\begin{document} 
\begin{titlepage}
\renewcommand{\thefootnote}{\fnsymbol{footnote}}
\begin{flushright}
ZU-TH 28/15
\end{flushright}
\vspace*{2cm}

\begin{center}
{\Large \bf The $q_T$ subtraction method\\[0.2cm] for top quark production at hadron colliders}
\end{center}

\par \vspace{2mm}
\begin{center}
{\bf Roberto Bonciani$^{(a)}$},
{\bf Stefano Catani$^{(b)}$},\\[0.2cm]
{\bf Massimiliano Grazzini$^{(c)}$\footnote{On leave of absence from INFN, Sezione di Firenze, Sesto Fiorentino, Florence, Italy.}},
{\bf Hayk Sargsyan$^{(c)}$} and {\bf Alessandro Torre$^{(c)}$}
\vspace{5mm}

$^{(a)}$ Dipartimento di Fisica, Universit\`a di Roma ``La Sapienza'' and\\
INFN, Sezione di Roma, I-00185, Roma, Italy

$^{(b)}$ INFN, Sezione di Firenze and Dipartimento di Fisica e Astronomia,\\
Universit\`a di Firenze, I-50019 Sesto Fiorentino, Florence, Italy

$^{(c)}$Physik-Institut, Universit\"at Z\"urich, CH-8057 Z\"urich, Switzerland

\vspace{5mm}

\end{center}

\par \vspace{2mm}
\begin{center} {\large \bf Abstract} \end{center}
\begin{quote}
\pretolerance 10000

We consider QCD radiative corrections to top-quark pair production at hadron colliders.
We use the $q_T$ subtraction
formalism to perform a fully-differential computation for this process.
Our calculation is accurate up to the next-to-leading order in QCD perturbation
theory and it includes all the flavour off-diagonal partonic channels
%the $q({\bar q})g$ and $q({\bar q})q^\prime$ channels 
at the next-to-next-to-leading order.
We present a comparison of our numerical results with those obtained 
with the publicly available numerical programs {\sc MCFM} and {\sc Top++}.

\end{quote}

\vspace*{\fill}
\begin{flushleft}
August 2015

\end{flushleft}
\end{titlepage}

\setcounter{footnote}{1}
\renewcommand{\thefootnote}{\fnsymbol{footnote}}

The top quark ($t$) has a special role 
\cite{Agashe:2014kda} in elementary particle physics. Being
the heaviest known fundamental constituent, with a mass of about 173.3~GeV 
\cite{ATLAS:2014wva}, it
couples strongly to the Higgs boson and it is crucial to the hierarchy problem.
Within the Standard Model (SM)
the main source of top-quark events in collisions at hadron
colliders is top-quark pair production. Many New Physics (NP) models predict
the existence of top partners with masses close to the electroweak
symmetry breaking scale, which exhibit similar properties as the top quark
and can decay into it.
Studying the production of $t{\bar t}$ pairs at hadron colliders
can not only shed light on the nature of the electroweak-symmetry breaking but
it also provides information on the backgrounds of many NP models.

The theoretical efforts for obtaining precision predictions for top-quark pair 
production at hadron colliders started almost three decades ago with 
the calculation of the next-to-leading order (NLO) QCD corrections 
to the total cross section 
\cite{Nason:1987xz, Beenakker:1988bq, Nason:1989zy}
and kinematical distributions \cite{Mangano:1991jk}
for this production process.
%% ADDED
The NLO calculations of the total cross section of
Refs.~\cite{Nason:1987xz, Beenakker:1988bq, Nason:1989zy} were carried out numerically. The expressions in
analytic form of the total partonic cross section\footnote{A parametrization
\cite{Langenfeld:2009wd}
of this analytic NLO result
is implemented in the numerical programs
{\sc Hathor} \cite{Aliev:2010zk}
and {\sc Top++} \cite{Czakon:2011xx}.}
at NLO were obtained in Ref.~\cite{Czakon:2008ii}.
Recently the calculation of the next-to-next-to-leading order (NNLO) QCD 
corrections to the $t{\bar t}$ total cross section was completed
%\cite{Baernreuther:2012ws, Czakon:2012zr, Czakon:2012pz, Czakon:2013goa}.
\cite{Baernreuther:2012ws}.
Besides the total cross section,  differential cross sections and more general
kinematical distributions are of 
great importance for precision studies. 
For instance, the $t{\bar t}$ (forward--backward and charge) asymmetry has
received much attention in recent years (see, e.g., Ref.~\cite{Kuehn:2014rla}).
The $t{\bar t}$ asymmetry, which is non vanishing starting from the NLO level \cite{Kuhn:1998kw},
has recently been computed up to the NNLO level \cite{Czakon:2014xsa}.
Other NNLO results on differential distributions 
are starting to appear \cite{Czakon:2015pga, Abelof:2014jna, Czakontalk}.

This Letter is devoted to the NNLO (and NLO) QCD calculation of 
$t{\bar t}$ production. In particular, we present the results of the first NNLO
application of the $q_T$ subtraction formalism~\cite{Catani:2007vq} to the process
of $t{\bar t}$ production in hadron collisions.

At the partonic level, the NNLO calculation of $t{\bar t}$ production requires 
the evaluation of tree-level contributions with two additional
unresolved partons in the final state, 
of one-loop contributions with one unresolved parton and 
of purely virtual contributions.
%of two-loop virtual corrections. 
%of virtual corrections from two-loop scattering amplitudes and the square of 
%one-loop scattering amplitudes.
The required tree-level and one-loop scattering amplitudes
are known and they are the same amplitudes that enter the NLO calculation of 
$t{\bar t}+{\rm jet}$ \cite{Dittmaier:2007wz}, the associated production of a 
$t{\bar t}$ pair and one jet.
The purely virtual contributions depend on the two-loop scattering amplitudes
and on the square of one-loop scattering amplitudes.
The two-loop amplitude for $t{\bar t}$ production is partly known in analytic
form \cite{Bonciani:2009nb} and its complete computation 
has been carried out numerically  \cite{Czakon:2008zk}.
%only some of its contributions are analytically available \cite{Bonciani:2009nb},
%while the complete computation has been carried out numerically 
%\cite{Czakon:2008zk}.
The square of one-loop scattering amplitudes is known \cite{Korner:2008bn}.

The implementation of the various scattering amplitudes in a complete 
NNLO calculation at the fully differential (exclusive) level is
a highly non-trivial task because of the presence of infrared (IR) 
divergences at intermediate stages of the calculation. 
In particular, these divergences do not permit a straightforward 
implementation of numerical techniques.
Various methods have been proposed and used to overcome these difficulties 
at the NNLO level.
The formalisms of {\em antenna} subtraction 
\cite{Kosower:1997zr, GehrmannDeRidder:2005cm,Daleo:2006xa,Currie:2013vh}
and {\em colourful} subtraction \cite{Somogyi:2005xz, DelDuca:2015zqa}
are more related to NNLO extensions of established 
NLO formulations \cite{Frixione:1995ms, Catani:1996jh, Catani:2002hc} 
of the subtraction method.
The {\em Stripper} formalism \cite{Czakon:2010td,Czakon:2011ve,Czakon:2014oma} 
is a combination of the subtraction method with
numerical techniques based on 
{\em sector decomposition} \cite{Anastasiou:2003gr,Binoth:2000ps}.
Variants of the subtraction methods are the 
$q_T$ subtraction formalism \cite{Catani:2007vq} and the recently proposed 
$N$-{\em jettiness} subtraction 
\cite{Boughezal:2015dva,Boughezal:2015eha,Gaunt:2015pea}.

The NNLO computations in Refs.~\cite{Baernreuther:2012ws,Czakon:2014xsa}
%Refs.~\cite{Baernreuther:2012ws, Czakon:2012zr, Czakon:2012pz, Czakon:2013goa,
%Czakon:2014xsa} 
for $t{\bar t}$ production have been performed by using the
Stripper method \cite{Czakon:2010td}.
Parallely, an ongoing effort is being carried out by using the 
antenna subtraction method \cite{GehrmannDeRidder:2005cm, Abelof:2011jv}, 
which led to the NNLO fully differential
computation of $t{\bar t}$ production in the
$q{\bar q}$ channel \cite{Abelof:2014fza, Abelof:2014jna}
at leading colour and including the light-quark contributions.

The $q_T$ subtraction formalism~\cite{Catani:2007vq} is
a method to handle and cancel the IR divergences at the NLO and NNLO level.
The method has been successfully applied to
the fully differential computation of NNLO QCD corrections to several
hard-scattering processes
%\cite{Catani:2007vq,Catani:2009sm,Ferrera:2011bk,Catani:2011qz,
%Grazzini:2013bna,Ferrera:2014lca,Cascioli:2014yka,Gehrmann:2014fva,
%Grazzini:2015nwa}.
\cite{Catani:2007vq,Grazzini:2008tf, Catani:2009sm, qtappl}.
The method uses IR subtraction counterterms that are constructed by considering
and explicitly computing the transverse-momentum ($q_T$) distribution
of the produced final-state system in the limit $q_T \to 0$.
If the produced final-state system is composed of non-QCD (colourless) partons
(e.g., leptons, vector bosons or Higgs bosons), the behaviour of the 
$q_T$ distribution in the limit $q_T \to 0$ has a universal 
(process-independent) structure that is explicitly known up to the NNLO level
through the formalism of transverse-momentum resummation 
\cite{Catani:2013tia}. These results on transverse-momentum resummation
are sufficient to fully specify the $q_T$ subtraction  formalism for this entire
class of processes. Therefore,
up to now, the applications of the $q_T$ subtraction formalism have been 
limited to the production of colourless
high-mass systems in hadron collisions.
In this Letter we present first results on the application of the 
$q_T$ subtraction method to the NNLO computation of heavy-quark production
in hadron collisions.
To this purpose, we use the recent progress on transverse-momentum resummation
for heavy-quark production \cite{Zhu:2012ts,Li:2013mia,Catani:2014qha}.
We exploit the formulation of transverse-momentum resummation
in Ref.~\cite{Catani:2014qha} that includes the {\em complete} dependence on the
kinematics of the heavy-quark pair. This dependence and, in particular, the
complete control on the heavy-quark azimuthal correlations are essential (see
below) to extract all the NNLO counterterms of the $q_T$ subtraction method. 
Although the structure of transverse-momentum resummation for heavy-quark
production is fully worked out up to the NNLO level, the explicit NNLO results
for the hard-virtual factors \cite{Catani:2014qha} in the flavour diagonal
partonic channels $q{\bar q} \to t {\bar t}+X$ and $gg \to t {\bar t}+X$
($X$ denotes the unobserved inclusive final state)
are not yet known. Therefore, in the NNLO calculation of this paper
we present numerical results for all the flavour off-diagonal channels
$ab \to t {\bar t}+X$, with $ab= qg ({\bar q}g), qq ({\bar q}{\bar q}),
qq' ({\bar q}{\bar q}'), q{\bar q}' ({\bar q} q')$ ($q$ and $q'$ denote quarks
with different flavour).

The differential cross section $d{\sigma}^{t{\bar t}}$ for 
the inclusive production process $pp (p {\bar p})\to t{\bar t}+X$ is computable
by convoluting the corresponding partonic cross sections 
$d{\hat \sigma}^{t{\bar t}}_{ab}$ of the various partonic channels
with the parton distribution functions (PDFs) of the colliding hadrons $pp\,(p {\bar p})$.
According to the $q_T$ subtraction method~\cite{Catani:2007vq}, the (N)NLO
%differential 
partonic cross section $d{\hat \sigma}^{t{\bar t}}_{(N)NLO}$ 
%for the inclusive production 
%process $pp (p {\bar p})\to t{\bar t}+X$
can be written as
\begin{equation}
\label{eq:main}
d{\hat \sigma}^{t{\bar t}}_{(N)NLO}={\cal H}^{t{\bar t}}_{(N)NLO}\otimes d{\hat \sigma}^{t{\bar t}}_{LO}
+\left[ d{\hat \sigma}^{t{\bar t}+\rm{jet}}_{(N)LO}-
d{\hat \sigma}^{t{\bar t}, \, CT}_{(N)NLO}\right],
\end{equation}
where $d{\hat \sigma}^{t{\bar t}+\rm{jet}}_{(N)LO}$ is the $t{\bar t}$+jet cross 
section 
at (N)LO accuracy.
Applying Eq.~(\ref{eq:main}) at NLO,  the leading-order (LO) cross section
$d{\sigma}^{t{\bar t}+\rm{jet}}_{LO}$
can be directly obtained by integrating the corresponding tree-level scattering
amplitudes. Applying Eq.~(\ref{eq:main}) at NNLO,
$d{\sigma}^{t{\bar t}+\rm{jet}}_{NLO}$
can be evaluated by using any available NLO method 
(e.g., Refs.~\cite{Frixione:1995ms, Catani:1996jh, Catani:2002hc})
to handle and cancel the corresponding IR divergences.
Therefore, $d{\sigma}^{t{\bar t}+\rm{jet}}_{(N)LO}$ is IR finite {\em provided}
$q_T \neq 0$.

The square bracket term of Eq.~(\ref{eq:main}) is IR finite in the limit
$q_T \to 0$, but its individual contributions,
$d{\sigma}^{t{\bar t}+\rm{jet}}_{(N)LO}$ and
$d{\sigma}^{t{\bar t}, \, CT}_{(N)NLO}$, are separately divergent.
The IR subtraction counterterm $d{\sigma}^{t{\bar t}, \,CT}_{(N)NLO}$
is obtained from the (N)NLO perturbative expansion 
(see, e.g., Ref.~\cite{Bozzi:2005wk,Bozzi:2007pn})
of the resummation formula
of the logarithmically-enhanced
contributions to the $q_T$ distribution
of the $t{\bar t}$ pair \cite{Zhu:2012ts,Li:2013mia,Catani:2014qha}:
the explicit form of $d{\sigma}^{t{\bar t}, \,CT}_{(N)NLO}$ 
can be completely worked out up to NNLO accuracy. 
For example, at the NLO, the explicit expression of $d{\hat \sigma}^{t{\bar t}, \,CT}_{NLO}$ in the partonic channel 
%$ab$ 
$ab \to t {\bar t}+X$ is
\begin{equation}
\label{ctnlo}
d{\hat \sigma}^{t{\bar t}, \,CT}_{NLO\, ab}=\sum_{c=q,{\bar q},g}
\f{\as}{\pi}
\left(\Sigma^{(1)}_{c{\bar c}\ito ab}
+ \Sigma_{c{\bar c}\ito ab}^{(1)t{\bar t}-{\rm new}} \right) \f{dq_T^2}{M^2}
\otimes d{\hat \sigma}^{t{\bar t}}_{LO\, c{\bar c}} \;\;,
%+ \sum_{i,j=1}^{4}\Sigma_{c{\bar c}\ito ab}^{(1){\rm new}(ij)}\cdot d\sigma^{t{\bar t}\,(ij)}_{LO\, c{\bar c}}\right) .
\end{equation}
where $\as$ is the QCD coupling, $M$ is the invariant mass of the produced 
$t{\bar t}$ pair and $d{\hat \sigma}^{t{\bar t}}_{LO\, ab}$ is the LO partonic
cross section. 
The expression (\ref{ctnlo}) involves convolutions (which are denoted by the
symbol $\otimes$) with respect to the longitudinal-momentum fractions
$z_1$ and $z_2$ of the colliding partons $c$ and $\bar c$ in 
$d{\hat \sigma}^{t{\bar t}}_{LO\, c{\bar c}}$.
The integration variable $q_T$ in Eq.~(\ref{ctnlo}) corresponds,
in the limit $q_T \to 0$, to the transverse momentum
of the produced $t{\bar t}$ pair in the cross section 
$d{\sigma}^{t{\bar t}+\rm{jet}}_{LO}$ on the right-hand side of 
Eq.~(\ref{eq:main}).
The function $\Sigma^{(1)}_{c{\bar c}\ito ab}$ enters into the $q_T$ subtraction
method \cite{Catani:2007vq} for hard-scattering production of a generic
final-state system. Its explicit form is \cite{Bozzi:2005wk,Bozzi:2007pn}
\begin{align}
\label{sigma1}
\Sigma^{(1)}_{c{\bar c}\ito ab}(z_1,z_2;q_T/M)=&-\f{1}{2}A_c^{(1)}\delta_{ca}\delta_{{\bar c}b}\delta(1-z_1)\delta(1-z_2)\,{\tilde I}_2(q_T/M)-\Big[\delta_{ca}\delta_{{\bar c}b}\delta(1-z_1)\delta(1-z_2)\, B^{(1)}_c\nn\\
&+\delta_{ca}\delta(1-z_1)\,P_{{\bar c}b}^{(1)}(z_2)+\delta_{{\bar
c}b}\delta(1-z_2)\,P_{ca}^{(1)}(z_1)\Big]{\tilde I}_1(q_T/M) \;\;,
\end{align}
and it derives from the small-$q_T$
singular behavior of the $q_T$ cross section for the production of a colourless
system in the partonic $c{\bar c}$ production channel. The coefficients
$A_c^{(1)}$ and $B_c^{(1)}$ are the first-order resummation coefficients for transverse-momentum resummation ($A^{(1)}_q=C_F$, $A^{(1)}_g=C_A$, $B^{(1)}_q=-3/2\, C_F$, $B^{(1)}_g=-(11/6\, C_A-n_F/3)$).
The functions $P_{ab}^{(1)}(z)$ are the lowest-order DGLAP kernels 
(the overall normalizazion is specified according to the notation in Eq.~(41) of
Ref.~\cite{Bozzi:2005wk}).
%and the symbol $\otimes$ denotes a standard convolution.
The functions ${\tilde I}_k(q_T/M)$ $(k=1,2)$, which appear in 
%Eqs.~(\ref{sigma1}) and (\ref{sigma1new}) 
Eq.~(\ref{sigma1}),
encapsulate the singular behavior at small $q_T$, and they are
explicitly given in Appendix~B of Ref.~\cite{Bozzi:2005wk}.
The other function 
$\Sigma_{c{\bar c}\ito ab}^{(1)t{\bar t}-{\rm new}}$ 
in the round-bracket factor of Eq.~(\ref{ctnlo}) is due to soft radiation and it
is an additional term that is specific of the $q_T$ subtraction method for the
case of heavy-quark pair production.
This function reads
\begin{equation}
\label{sigma1new}
\Sigma^{(1)t{\bar t}-{\rm new}}_{c{\bar c}\ito ab}(z_1,z_2;q_T/M)
=-\delta_{ca}\delta_{{\bar c}b} \delta(1-z_1)\delta(1-z_2)
\frac{\langle {\cal M}_{c{\bar c}\to t{\bar t}}|\left({\bf \Gamma}^{(1)}_t+{\bf \Gamma}_t^{(1)\dagger}\right)|{\cal M}_{c{\bar c}\to t{\bar t}}\rangle}{|{\cal M}_{c{\bar c}\to t{\bar t}}|^2} \, {\tilde I}_1(q_T/M)\, ,
\end{equation}
where ${\bf \Gamma}^{(1)}_t$
is the first-order term of the soft anomalous dimension for transverse-momentum 
resummation in heavy-quark production and its explicit expression 
is given in Eq.~(33) of Ref.~\cite{Catani:2014qha}.
This soft anomalous dimension is a colour space matrix that acts onto the colour
indices of the four partons $\{c, {\bar c}, t, {\bar t} \}$ in the Born level
scattering amplitude 
$|{\cal M}_{c{\bar c}\to t{\bar t}}\rangle$
of the partonic process $c{\bar c}\to t{\bar t}$.
The colour space notation is
specified in Ref.~\cite{Catani:2014qha} and, in particular, 
$|{\cal M}_{c{\bar c}\to t{\bar t}}|^2 = 
\langle {\cal M}_{c{\bar c}\to t{\bar t}}|{\cal M}_{c{\bar c}\to t{\bar t}}\rangle$
denotes the colour summed square amplitude that contributes to 
$d{\hat \sigma}^{t{\bar t}}_{LO\, c{\bar c}}$, whereas the factor
$\langle {\cal M}_{c{\bar c}\to t{\bar t}}|({\bf \Gamma}^{(1)}_t+{\bf \Gamma}_t^{(1)\dagger})|{\cal M}_{c{\bar c}\to t{\bar t}}\rangle$
embodies colour correlation terms with a definite kinematical dependence.

The first-order
%hard-virtual
hard-collinear (IR finite) counterterms 
%coefficients 
${\cal H}^{t{\bar t}}_{NLO}$ are also completely known 
\cite{Zhu:2012ts,Li:2013mia,Catani:2014qha} for all the partonic channels.
The second-order 
%hard-virtual 
(IR finite) counterterms
%coefficients 
${\cal H}^{t{\bar t}}_{NNLO}$ are not yet fully known.
However, ${\cal H}^{t{\bar t}}_{NNLO}$ can be explicitly determined
for all the flavour off-diagonal partonic channels.
%($q({\bar q})g\to t{\bar t}$, $q({\bar q})q^\prime\to t{\bar t}$....) 
In these off-diagonal channels, ${\cal H}^{t{\bar t}}_{NNLO}$ embodies
process-dependent and process-independent contributions. The 
process-dependent contributions to ${\cal H}^{t{\bar t}}_{NNLO}$
derive from the knowledge of the one-loop virtual amplitudes of the partonic
processes $q{\bar q} \to t{\bar t}$ and $gg\to t{\bar t}$,
and from the explicit results on the NLO {\em azimuthal correlation} terms
in the transverse-momentum resummation formalism \cite{Catani:2014qha}
(see, in particular, Eq.~(25) in Ref.~\cite{Catani:2014qha} and accompanying
comments). The process-independent contributions to 
${\cal H}^{t{\bar t}}_{NNLO}$ are analogous to those that contribute to Higgs
boson \cite{Catani:2007vq} and vector boson \cite{Catani:2009sm} production,
and they are explicitly known 
%\cite{Catani:2011kr,Catani:2012qa}.
\cite{Catani:2011kr,Catani:2013tia,Gehrmann:2012ze}.

Having discussed the content of Eq.~(\ref{eq:main}), 
we are in a position to apply it to $t{\bar t}$ production and to 
obtain the complete NLO results 
plus the NNLO corrections in all the flavour off-diagonal partonic channels. 
Our NLO implementation of the calculation has the main purpose of illustrating
the applicability of the $q_T$ subtraction method  to heavy-quark production
and, in particular, of cross-checking the $q_T$ subtraction methodology by
numerical comparisons with NLO calculations performed by using more
established NLO methods. Our NNLO results on $t{\bar t}$ production 
represent a first step (due to the missing flavour diagonal partonic channels)
towards the complete NNLO calculation for this production process.
Up to NLO accuracy our numerical implementation is based on
the scattering amplitudes and phase space generation of the {\sc MCFM}
program \cite{mcfm}, 
suitably modified for $q_T$ subtraction along the
lines of the corresponding numerical programs for 
Higgs boson \cite{Catani:2007vq} and vector boson \cite{Catani:2009sm} 
production.
At NNLO accuracy the $t{\bar t}$+jet cross section is evaluated by using 
the {\sc Munich} code \cite{Kallweit:Munich}, which
provides a fully automated implementation of the 
%Catani--Seymour 
NLO dipole subtraction
formalism 
%\cite{Catani:1996jh,Catani:1996vz,Catani:2002hc}
\cite{Catani:1996jh,Catani:2002hc}
as well as an interface to the one-loop generator {\sc OpenLoops} 
\cite{Cascioli:2011va} to obtain all the required (spin- and colour-correlated) 
tree-level and one-loop amplitudes.
For the evaluation of tensor integrals we rely on the 
\textsc{Collier} library \cite{Denner:2014gla}, which is based on the 
Denner--Dittmaier reduction techniques \cite{Denner:2002ii} of tensor integrals
%\cite{Denner:2002ii,Denner:2005nn} 
and on the scalar integrals of Ref.~\cite{Denner:2010tr}.
In {\sc OpenLoops}
problematic phase space points are addressed with a rescue system 
that uses the quadruple-precision implementation of the 
OPP method 
in \textsc{CutTools}~\cite{Ossola:2007ax}
%, involving scalar integrals from \textsc{OneLOop}~\cite{vanHameren:2010cp}.
with scalar integrals from \textsc{OneLOop}~\cite{vanHameren:2010cp}.

%%====================================
\begin{figure}[th]
\centering
\hspace*{-0.8cm}
\subfigure[]{
\includegraphics[width=3.2in]{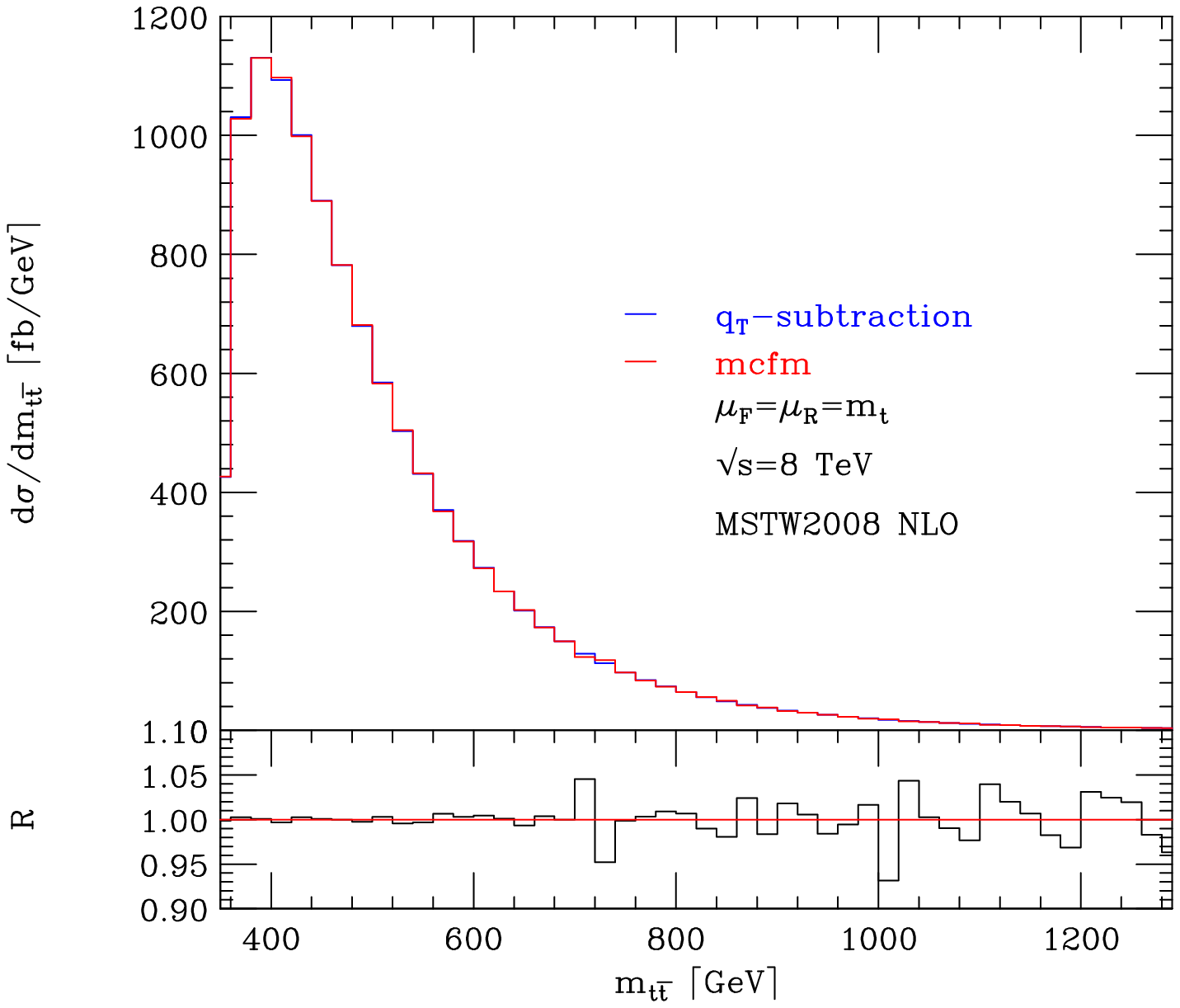}
}
\subfigure[]{
\includegraphics[width=3.2in]{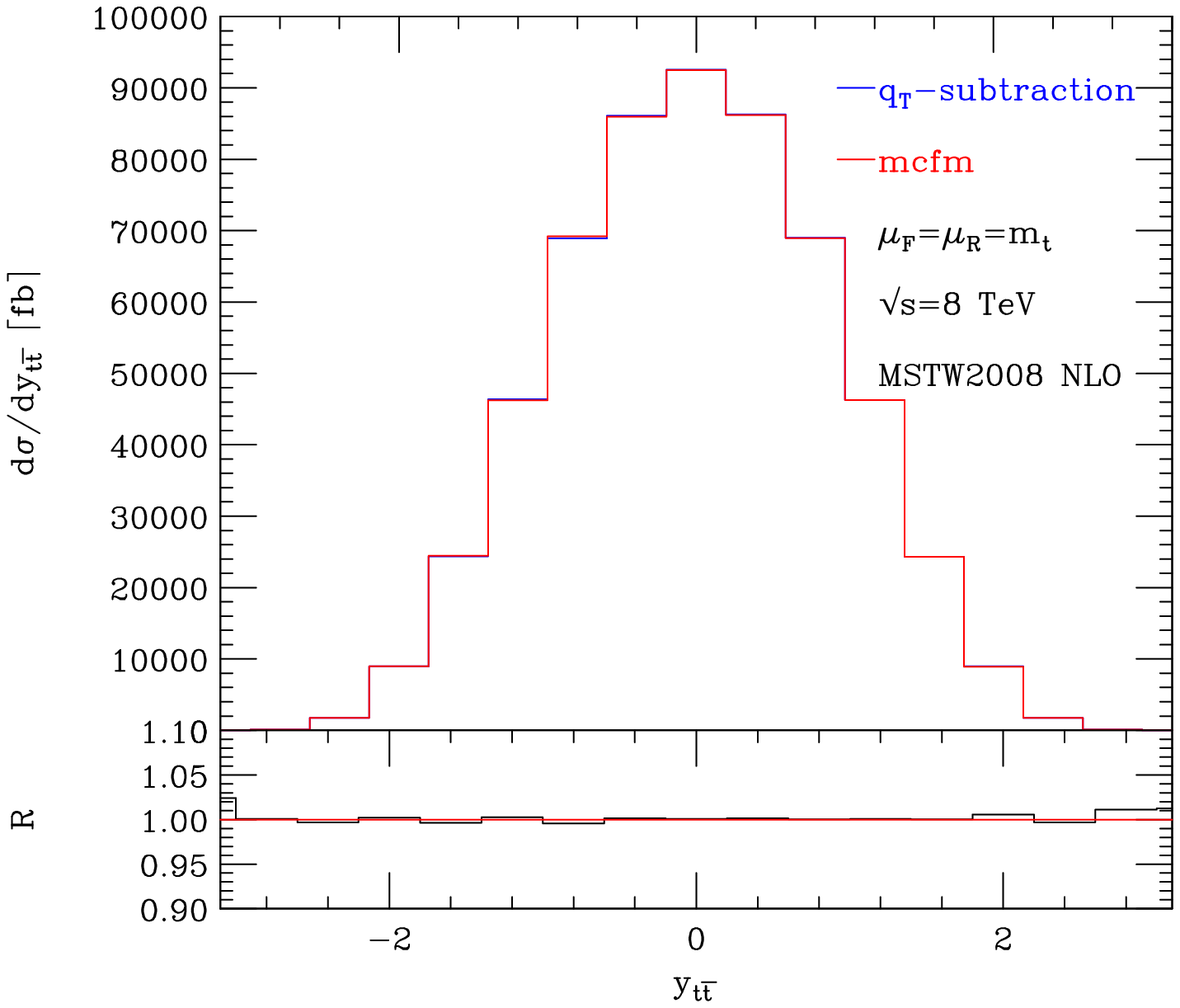}}
\caption{\label{fig:ttb}
The invariant mass (left) and rapidity (right) distributions of the $t\bar{t}$
pair at the LHC ($\sqrt{s}=8~\mathrm{TeV}$) computed at NLO accuracy.
Comparison of our results (blu) with the MCFM results (red). 
The lower panel presents the ratio of our results over the MCFM results.}
\end{figure}
%%====================================
%%====================================
\begin{figure}[th]
\centering
\hspace*{-0.8cm}
\subfigure[]{
\includegraphics[width=3.2in]{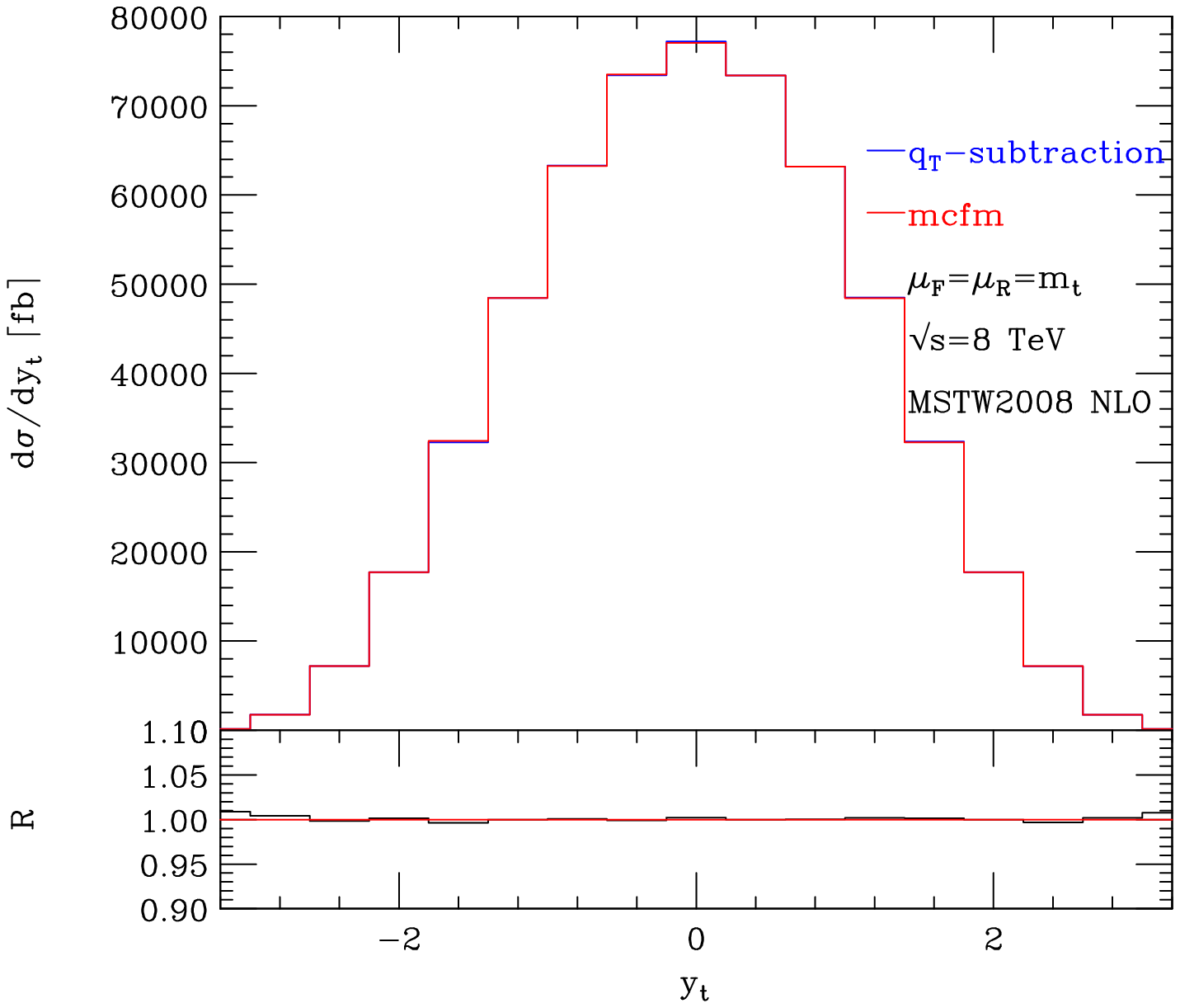}
}
\subfigure[]{
\includegraphics[width=3.2in]{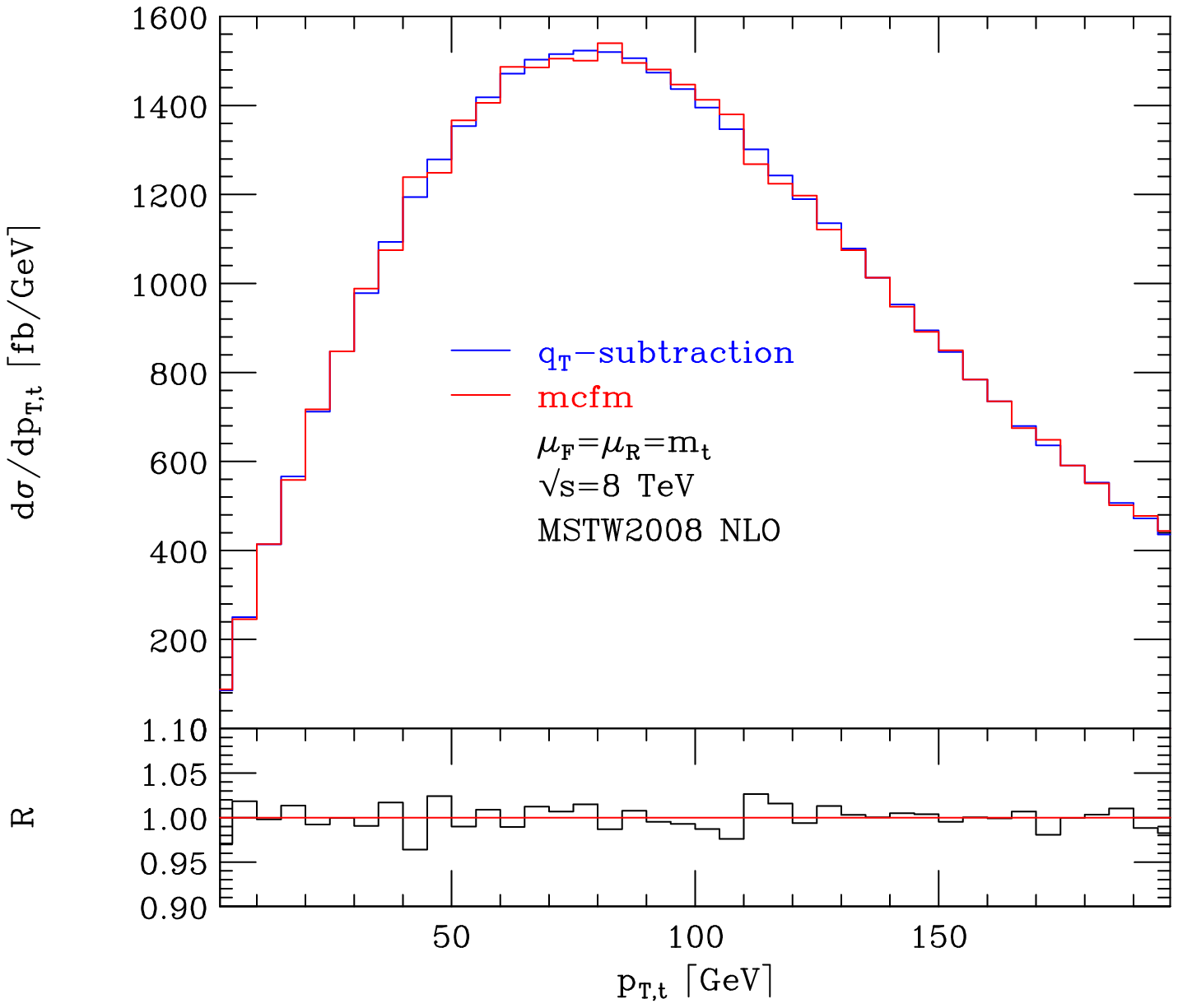}}
\caption{\label{fig:top} The rapidity (left) and transverse-momentum (right)
 distributions of the top quark at the LHC ($\sqrt{s}=8~\mathrm{TeV}$) computed
 at NLO accuracy. Comparison of our results (blu) with the MCFM results (red).
The lower panel presents the ratio of our results over the MCFM results.}
\end{figure}
%%====================================

We start the presentation of our results by considering $pp$ collisions at 
$\sqrt{s}=8~\mathrm{TeV}$.
We use the MSTW2008 \cite{Martin:2009iq} PDFs
%parton distribution functions (PDFs) 
with the QCD running coupling $\as$ evaluated
at each corresponding order
(i.e., we use $(n+1)$-loop $\as$ at N$^n$LO, with $n=1,2$). 
The pole mass of the top quark is $m_t=173.3$~GeV. The renormalization and
factorization scales, $\mu_R$ and $\mu_F$, are fixed at
$\mu_R=\mu_F=m_t$.

In Figs.~\ref{fig:ttb} and \ref{fig:top} we compare the NLO differential 
distributions obtained by using {\sc MCFM} (which implements the dipole
subtraction method \cite{Catani:1996jh,Catani:2002hc})
with those obtained by using our numerical program\footnote{These NLO results are obtained by using {\sc MCFM-v7.0} with about 9 hours of run on an Intel Xeon 2.4 Ghz, corresponding to a total of $2*10^6$ integral evaluations. Our NLO implementation of $q_T$ subtraction requires about $30$ times
higher statistics to obtain comparable results and the corresponding runtime is
a factor of 3 larger.}. 
In particular in Fig.~\ref{fig:ttb} we consider the 
invariant mass ($m_{t{\bar t}}$) distribution (left) and the rapidity 
($y_{t{\bar t}}$) distribution (right) 
of the $t{\bar t}$ pair.
In Fig.~\ref{fig:top} we consider the 
rapidity ($y_{t}$) distribution (left) and the transverse-momentum ($p_{T,t}$) 
distribution (right) of the top quark. We clearly see that the distributions 
obtained with $q_T$ subtraction are in excellent agreement with those 
obtained with {\sc MCFM}. 
We have checked that the agreement persists also for different choices 
of $\mu_R$ and $\mu_F$.

We now move to consider the NNLO contributions and, in particular,
we compute the total cross section for $t{\bar t}$ production.
In Table~\ref{tab:8tev} we report our results and we compare them 
with the corresponding results from the numerical program {\sc Top++}
\cite{Czakon:2011xx}, which implements the NNLO calculation of 
%Refs.~\cite{Baernreuther:2012ws, Czakon:2012zr, Czakon:2012pz, Czakon:2013goa}.
Ref.~\cite{Baernreuther:2012ws}.
Specifically, we report the complete NLO results and the ${\cal O}(\as^4)$
contributions to the NNLO cross section from the flavour off-diagonal partonic
channels $ab \to t {\bar t}+X$. The contribution from all the channels
with $ab= qg,{\bar q}g$ is labelled by the subscript $qg$, and the 
contribution from all the channels with 
$ab=qq, {\bar q}{\bar q}, qq', {\bar q}{\bar q}', q{\bar q}', {\bar q} q'$
is labelled by the subscript $q({\bar q})q^\prime$.
From Table~\ref{tab:8tev}
we see that the results obtained by using $q_T$ subtraction 
are in agreement with those from {\sc Top++}. 
We note that the numerical uncertainties of our ${\cal O}(\as^4)$ results are 
at the 2\% level. This is
due to the fact that in both the $qg$ and $q({\bar q})q^\prime$ channels 
at $\sqrt{s}=8$ TeV
there is a strong quantitative cancellation between the contributions
of the two terms in the right-hand side of Eq.~(\ref{eq:main})
(the term that is proportional to ${\cal H}^{t{\bar t}}_{NNLO}$ and the term
in the square bracket). 
The numerical uncertainties of our ${\cal O}(\as^4)$ calculation can be reduced
by considering different centre--of--mass energies. In particular, the numerical
cancellation that we have mentioned is reduced by decreasing the 
centre--of--mass energy of the collision.
%The quality of the comparison can be improved by 
%considering different centre-of-mass energies. 
We have computed the total cross section for $p{\bar p}$ collisions
at $\sqrt{s}=2$~TeV and we report the corresponding results in
Table~\ref{tab:2tev}. We note that the numerical agreement between our
calculation and the {\sc Top++} result is still satisfactory,
and the numerical uncertainties of our ${\cal O}(\as^4)$ results are
reduced below the 1\% level.
%smaller by about a factor of two.
Using {\sc Top++} we can compute
the complete NNLO result and we note that the flavour off-diagonal partonic
channels contribute to about 10\% of
the total result at ${\cal O}(\as^4)$
for both collision energies considered in Tables~\ref{tab:8tev} and
\ref{tab:2tev}.

%%%%%%%%%%%%%%%%%%%%%%%
\begin{table}[!ht]
\vspace*{6mm}
\begin{center}
\begin{tabular}{|c||c|c|c|}
\hline
Cross section [pb] & NLO & ${\cal O}(\as^4)|_{qg}$ & ${\cal O}(\as^4)|_{q({\bar q})q^\prime}$ \\ \hline \hline
$q_T$ subtraction  & $226.2(1)$ & $-2.25(5)$ & 0.151(3)\\ \hline
 {\sc Top++} &  $226.3$ &  $-2.253$& 0.148\\ \hline
\end{tabular}
\end{center}
\vspace*{-2mm}
\caption{Total cross sections for $t{\bar t}$ production.
NLO and (partial) NNLO results from $q_T$ subtraction compared 
with the corresponding results from {\sc Top++} for $pp$ collisions at 
$\sqrt{s}=8$~TeV.}
\label{tab:8tev}
\end{table}
%%%%%%%%%%%%%%%%%%%%%%%

%%%%%%%%%%%%%%%%%%%%%%%
\begin{table}[!ht]
\vspace*{6mm}
\begin{center}
\begin{tabular}{|c||c|c|c|}
\hline
Cross section [fb] & NLO & ${\cal O}(\as^4)|_{qg}$ & ${\cal O}(\as^4)|_{q({\bar q})q^\prime}$\\ \hline \hline
$q_T$ subtraction  & $7083(3)$ & $-61.5(5)$& 1.33(1)\\ \hline
 {\sc Top++} &  $7086$ &  $-61.53$& 1.33\\ \hline
\end{tabular}
\end{center}
\vspace*{-2mm}
\caption{Total cross sections for $t{\bar t}$ production.
NLO and (partial) NNLO results from $q_T$ subtraction compared with the corresponding results from {\sc Top++} for $p{\bar p}$ collisions at $\sqrt{s}=2$ TeV.}
\label{tab:2tev}
\end{table}
%%%%%%%%%%%%%%%%%%%%%%%

We have presented the first application of the $q_T$ subtraction method
to top-quark pair production at hadron colliders. Our implementation is
based on the formulation of transverse-momentum resummation
for heavy-quark production of Ref.~\cite{Catani:2014qha}, which includes
the complete dependence on the
kinematics of the heavy-quark pair.
Our computation is accurate at NLO in
QCD perturbation theory and it includes all the off-diagonal partonic channels 
at NNLO accuracy.
At NLO we have compared our results for various kinematical distributions
with those obtained by using the {\sc MCFM} program, and we find good agreement.
At NNLO our results for the $t{\bar t}$ total cross section agree
with the corresponding results obtained by using the {\sc Top++} program. 
The extension of
our NNLO computation to include the missing
$q{\bar q} \to t{\bar t}+X$ and $gg\to t{\bar t}+X$ channels 
requires the evaluation
of the second-order hard-collinear functions ${\cal H}^{t{\bar t}}_{NNLO}$
\cite{Catani:2014qha},
and an implementation of the two-loop virtual amplitudes, which, at present,
are known only in numerical form \cite{Czakon:2008zk}.

The computation that we have performed in this paper
can straightforwardly be extended to the production
of massive-quark pairs of different flavour (e.g. bottom-quark pair). The extension of the method to production processes with massless coloured particles in the final state (e.g. inclusive dijet production) is definitely non trivial and it would require additional theoretical advancements.

\paragraph{Acknowledgements.} We are grateful to Stefan Kallweit for his help with the {\sc Munich} code. This research was supported in part by the Swiss National Science Foundation (SNF) under contract 200021-156585 and by the Research Executive Agency (REA) of the 
European Union under the Grant Agreement number PITN-GA-2012-316704 ({\it Higgstools}).
The work of RB was partly supported by the European Community Seventh Framework
Programme FP7/2007-2013, under grant agreement N.302997.
We acknowledge the hospitality and partial support 
of the Galileo Galilei Institute (GGI) in Florence,
where part of this work has been carried out during the Workshop
{\it Prospects and Precision at the Large Hadron Collider at 14~TeV}.


\begin{thebibliography}{99}
\bibitem{Kallweit:Munich}
S.~Kallweit,
{\sc Munich} is the abbreviation of ``MUlti-chaNnel Integrator at Swiss~(CH) precision''---an automated parton level NLO 
generator. 
In preparation.
\end{thebibliography}
\end{document}